# Optical frequency synthesis and measurement using fibre-based femtosecond lasers

Gesine Grosche, Burghard Lipphardt und Harald Schnatz

Physikalisch-Technische Bundesanstalt, Bundesallee 100, D-38116 Braunschweig, Germany

**Abstract:** We report the synthesis and measurement of an ultra-precise and extremely stable optical frequency in the telecommunications window around 1543 nm. Using a fibre-based femtosecond frequency comb we have phase-stabilised a fibre laser at 194 THz to an optical frequency standard at 344 THz, thus transferring the properties of the optical frequency standard to another spectral region. Relative to the optical frequency standard, the synthesised frequency at 194 THz is determined to within 1 mHz and its fractional frequency instability is measured to be less than $2 \times 10^{-15}$ at 1 s, reaching $5 \times 10^{-18}$ after 8000 s. We also measured the synthesised frequency against a caesium fountain clock: here the frequency comparison itself contributes less than 4 mHz ($2 \times 10^{-17}$) to the uncertainty. Our results confirm the suitability of fibre based frequency comb technology for precision measurements and frequency synthesis, and enable long-distance comparison of optical clocks by using optical fibres to transmit the frequency information.

## 1. INTRODUCTION

Frequency combs based on mode-locked femtosecond lasers have revolutionised the field of frequency metrology and precision spectroscopy [1]. Linking the optical domain to the radio frequency domain, they have enabled measurements of optical transitions with a precision below 1 part in $10^{15}$ [2]. Such measurements are ultimately limited by the frequency instability and uncertainty of the microwave reference realizing the SI-second, such as a caesium fountain clock [3]. As optical frequency standards begin to outperform their microwave counterparts [4, 5], future measurements are likely to be direct comparisons between different optical frequency standards. This means that the *ratio* of two optical frequencies needs to be measured – typically over several hours, aiming for a relative uncertainty well below $10^{-16}$.

Introducing the *transfer oscillator* concept [6,7], Stenger *et al.* measured the known ratio of 2 between the fundamental, at 1064 nm, and the second harmonic optical frequency, at 532 nm, of a Nd:YAG laser with an uncertainty below $10^{-18}$ using a femto-second comb generator based on a Ti:Sapphire laser [7]; with the same technique they obtained a relative uncertainty – limited by the optical reference – approaching $10^{-14}$ for the ratio measurements of different optical sources [6]. We now report a relative uncertainty approaching $10^{-18}$ for a long-term ratio measurement of two optical frequencies originating from a diode laser operating at 871 nm (344 THz) and a fibre laser emitting near 1543 nm (194 THz).

In many precision physics experiments optical oscillators with excellent short-term instability down to $10^{-15}$ (1s) are essential, and intricate constructions have been devised to this end [8, 9]. At the same time, scientific and industrial applications require precise reference wave-lengths or optical frequency standards: for optical telecommunications, these have been realised through painstaking stabilisation of diode lasers to Doppler-free overtone transitions of acetylene [10]; for this method a frequency instability around $10^{-13}$, and an uncertainty within $10^{-11}$, have recently been reported [11, 12]. A possible alternative is *optical frequency synthesis* using femtosecond frequency comb (fs-comb) technology. Starting from a microwave oscillator, such synthesis has yielded an optical frequency instability of about $2 \times 10^{-14}$ at measurement times near 1 s [13]. Expecting superior results, we chose to study the fs-comb based synthesis of an optical frequency starting from a suitable *optical* reference. Using a fibre-based fs-comb we synthesised an optical target frequency at 194 THz (at 1543 nm, inside the telecommunication band) from a reference laser at 344 THz (871 nm) which is part of the Yb$^+$ optical frequency standard [5]. We thus transferred its properties to 194 THz, achieving a frequency uncertainty below 0.4 Hz, or relative uncertainty of $2 \times 10^{-15}$. Such frequency synthesis can generate multiple optical frequencies for spectroscopy or other applications from just one super-stable reference laser; working at 1.5 μm, an ultra-precise optical frequency could be distributed to potential users using optical telecommunication fibre networks.

To date, most frequency comb systems are based on Ti:Sapphire femtosecond lasers, and many experiments have established that ultra-precise frequency measurements with a relative uncertainty below $10^{-18}$ are possible with this technology [14]. Fibre laser based fs-combs, operating at 1.5 μm [15, 16], now combine the field of frequency metrology with optical telecommunications. Their robust, compact set-up allows reliable long-term operation [17, 18], enabling a new class of experiments and applications.

Noise processes of fibre based fs-comb systems, and their limiting influence for the measuring accuracy, are still under investigation. For example, two fibre frequency combs measuring the same optical frequency were found to agree within a few parts in $10^{-16}$ [17], while Benkler *et al.* [19] derived an upper limit for the frequency instability of $1.4 \times 10^{-14}$ (τ Hz)$^{-1/2}$ when *correlated* noise contributions are suppressed by the transfer oscillator method. More recently, Swann *et al.* [20] investigated the comb line-width of self-referenced, fibre-laser-based frequency combs by measuring the heterodyne beat signal between two independent frequency combs that are phase locked to a common cw optical reference. The results show instrument-limited, sub-Hertz relative line-widths of the optical comb lines across the comb spectrum [20, 21]. A careful analysis of the noise contributions in fibre combs can be found in [22], showing that there is no reason that the fibre-laser frequency comb cannot exhibit very low residual phase noise. Our present measurements confirm that noise processes do not limit frequency measurements down to a relative accuracy of $5 \times 10^{-18}$, well below the systematic uncertainty projected for the best optical frequency standards.

Here we report on one simple experiment converting the stability of a master oscillator to a single frequency source at 1.5 μm using a fibre based frequency comb generator and converting it back using a second fibre fs-comb. We show how this one experiment addresses each of the issues below:

- synthesis of ultra-stable optical frequencies, and realisation of the meter in the important telecommunication window near 1.5 μm, enabling the distribution of high precision optical frequencies to experimentalists, and remote comparison of optical frequency standards using standard telecom fibre

- the long-term measurement of optical frequency ratios, and testing of the comb properties including the achievable measurement accuracy using fibre based femto-second combs.

## 2. EXPERIMENTAL SET-UP

An overview of the experiment is shown in Figure 1: a slave laser is phase-stabilised to a master laser via the first femtosecond comb, and the absolute frequencies, as well as the ratio of the optical frequencies, are measured with the first comb and independently with a second femtosecond comb.

The Yb$^+$ master laser at 871 nm ($\nu_{L1} \approx 344$ THz) is an extended cavity diode laser (ECDL) stabilised to a high-finesse cavity by the Pound-Drever-Hall technique. Its frequency-doubled output is used to interrogate the Yb$^+$ ion optical clock transition at 436 nm. More details can be found in [5]. The linewidth of this stabilized laser was estimated to be <5 Hz. A frequency comparison with the spectroscopy laser of the Ca optical frequency standard [9, 23] confirmed this result. Throughout this paper the master laser serves as optical frequency reference for our experiments.

The slave laser is a free-running, commercially available distributed-feed-back fibre laser (Koheras Adjustik) operating at 1543 nm ($\nu_{L2} \approx 194$ THz). The laser has a free-running line-width below 4 kHz; its frequency can be tuned using a piezo-electric element, with a tuning coefficient of approximately 15 MHz/V and a bandwidth of almost 10 kHz.

To investigate the fiber combs potential and limitations with respect to accurate frequency measurements we used two independent frequency comb generators (labelled comb1, comb2 in Figure 1) to search for possible small systematic errors. Both femtosecond combs interlink the different spectral regions of the master laser $\nu_{L1}$ and the slave laser $\nu_{L2}$ and are used for two independent measurements of the *same* optical frequencies.

Each comb system incorporates its own passively mode-locked Er:doped femto-second fibre laser. While the stretched-pulse additive pulse mode-locking mechanism [24] is common to both systems, they follow different designs and differ fundamentally in many details, which are described in [15, 17]. Both systems are equipped with optical outputs at 1.5 μm and additional amplifier branches with subsequent spectral broadening and frequency doubling.

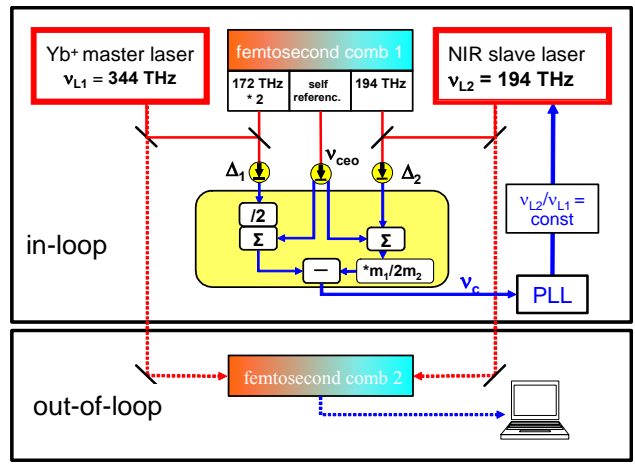

**Figure 1:** The set-up comprises an optical frequency standard $\nu_{L1}$ at 871 nm, a slave laser $\nu_{L2}$ at 1543 nm and two fs-frequency combs. While comb1 is used to phase-stabilize the slave laser to the master laser, comb2 is used to analyze the stability of the slave laser. The yellow box schematically shows the generation of the control signal for the laser stabilization. Similar electronics is used in the out-of-loop setup to utilize the transfer concept for the second frequency comb (for details see text).

The absolute frequency $\nu_m$ of the m-th comb-line of the fs-comb is determined by the pulse repetition frequency $f_{rep}$ and the carrier offset frequency $\nu_{ceo}$ according to

$$\nu_m(t) = m \cdot f_{rep}(t) + \nu_{ceo}(t) \qquad (1).$$

The beat signals $\Delta_1$, $\Delta_2$ of the two cw-lasers $\nu_{L1}$ and $\nu_{L2}$ with the comb line nearest to it are given by

$$\Delta_i(t) = \nu_{Li}(t) - \nu_m(t) \qquad i=1, 2 \quad (2).$$

In ref. [6], Telle *et al.* showed that by a deliberate choice of mixing frequencies the frequency noise introduced by the mode-locked laser drops out of the measurement process and it is possible to compare two narrow optical atomic lines without requiring the optical frequency comb to be highly stabilized. The advantage of this frequency transfer method is that it allows to transfer the frequency stability of the master laser to a very different optical frequency without adding additional noise from the comb.

Applying this concept, we derive a signal $\nu_C$ from $\Delta_1$, $\Delta_2$, and $\nu_{ceo}$ using standard analogue electronics (rf-mixers and filters) with an on-line bandwidth of 1 MHz. The generation of such an electronic signal is discussed in the following (see Fig. 1):

In a first step we generate the sum frequency of $\nu_{ceo}$ and $\Delta_i$ for both lasers taking into account that the beat signal $\Delta_1$ is derived from a beat with the frequency doubled output of the comb. This signal has to be divided by 2 before it is mixed with the carrier offset frequency $\nu_{ceo}$.

The second step compensates the different fluctuations of the comb modes at different optical frequencies. The electronic key element is a direct digital synthesizer (DDS) able to realize any rational numbers $((m_1/2)/m_2)$ with high precision. The DDS is used as a frequency divider for $(\nu_{ceo}+\Delta_2)$. Finally, taking the difference of the two signal paths we obtain

$$\nu_C = \left[ (\nu_{ceo} + \Delta_2) \div \frac{m_2}{m_1/2} - \left(\nu_{ceo} + \frac{\Delta_1}{2}\right) \right] \quad (3)$$

In terms of the optical frequencies involved, this beat can be expressed as

$$2\nu_C = \nu_{L2} \times \frac{m_1}{m_2} - \nu_{L1} \quad (4),$$

where $\nu_C$ can be considered as a beat note between the fixed master $\nu_{L1}$ and the tuneable slave laser $\nu_{L2}$.

We refer to this signal as *transfer beat* throughout the following text.

The transfer beat derived from comb1 is used to stabilise the fibre laser at 1543 nm to the 871 nm ECDL and is considered as in-loop signal in the following analysis. $\nu_C(t)$ is phase-stabilised to a chosen radio frequency $\nu_{ref}$ (in our case near 40 MHz). The error signal $\nu_C(t)-\nu_{ref}$ is applied to the tuning input of the slave laser in a feedback loop.

Using signals derived from comb2, a second electronic setup generates an independent transfer beat that corresponds to an out-of-loop measurement of the synthesised optical frequency $\nu_{L2}$.

Our setup allows to relate the frequency of the slave laser to that of a Cs fountain and an optical frequency standard simultaneously. While absolute frequency measurements (in SI Hertz) of optical frequency standards (see Sect. 3.1) are limited by the short term stability of the low frequency reference, measuring the ratio of two *optical* frequencies overcomes these limits as it takes advantage of the superior stability of an optical frequency standard (see Sect. 3.2).

This ratio

$$\frac{\nu_{L2}}{\nu_{L1}} = \frac{m_2}{m_1}\left\{1 + 2\frac{\nu_C}{\nu_{L1}}\right\} \quad (5),$$

is not referred to an absolute frequency and just mirrors the stability of the worse optical frequency standard.

This frequency $\nu_C$ is a measure for the (small) deviation of the optical signal's frequency ratio $\nu_{L2}/\nu_{L1}$ from $m_2/m_1$. Since $\nu_C \ll \nu_{L1}, \nu_{L2}$, the requirements on the radio frequency reference are in general not demanding: it is sufficient to refer $\nu_C$ to a caesium-clock controlled hydrogen maser.

All individual beat frequencies are pre-processed by tracking oscillators fast enough to follow the frequency fluctuations of the combs and lasers. All rf-frequencies are synthesised from a caesium fountain clock (CSF1). Even though not necessary, we have slowly locked the offset frequency $\nu_{ceo}$ and the pulse repetition frequency $f_{rep}$ to stable radio frequencies and thus kept the beat signals within the pass band of our fixed frequency filters. These pre-filters typically have a bandwidth of 5 MHz.

For data analysis and processing we used a high performance rf-spectrum analyzer and a multi channel accumulating counter with synchronous readout and zero dead-time, referenced to CSF1.

With this counter we recorded the frequencies $f_{rep}(t)$, $\nu_{ceo}(t)$, $\Delta_1(t)$, $\Delta_2(t)$ of each fs-comb and $\nu_C(t)$ with a gate time of 1 s.

Using equations (1) and (2) we then computed:

- the absolute frequencies $\nu_{L1}$ and $\nu_{L2}$, as measured with comb1
- the absolute frequencies $\nu_{L1}$ and $\nu_{L2}$, as measured with comb2
- the difference frequency between the combs, $\Delta_{NIR}(t) = \nu_{L2}(comb2,t) - \nu_{L2}(comb1,t)$,

Using equations (3) to (5) we also calculated:

- the frequency ratio $R_1 = \nu_{L2}/\nu_{L1}$ measured with comb1 (in-loop)
- the frequency ratio $R_2 = \nu_{L2}/\nu_{L1}$ measured with comb2 (out-of-loop)
- the difference of the frequency ratios $\Delta_R(t) = R_2 - R_1$

## 3. RESULTS

As a first application, we measured the stabilised optical frequency $\nu_{L2}$ at 194 THz (near 1543 nm) emitted by the NIR fibre laser, versus a caesium fountain as primary frequency standard (CSF1), and versus the optical frequency $\nu_{L1}$ at 344 THz (near 871 nm) of the Yb master laser over several days during measurement campaigns between July 2005 and June 2006. Long-term operation of the complete measurement set-up shown in Figure 1 was achieved with a duty-cycle over 95%: with a single fibre-based frequency comb we obtained continuous optical frequency measurements lasting for several days.

### 3.1. Absolute frequency measurement: measuring the synthesised optical frequency versus a caesium fountain

Figure 2 (red circles) shows the fractional instability, as measured by the Allan standard deviation (ADEV), of the NIR absolute frequency measurements for a continuous data-set lasting 42 h. During the same campaign, the frequency of the $^{171}Yb^+$ optical frequency standard was recorded (Fig. 2 green open triangles), as discussed elsewhere [25, 26]. The ADEV data for Yb and NIR are very similar, the fractional instabilities are approximately $10^{-13}$ at 1 s, and decrease as $\tau^{-1/2}$ for $\tau > 100$s. These data lie just below results obtained by Oskay et al. [2] (shown as top trace, continuous line in Fig. 2), who measured the frequency of an $Hg^+$ optical standard versus a caesium fountain.

We use a similar data analysis to that proposed by Oskay et al.: approximating our ADEV data as $2 \times 10^{-13}$ $(\tau\,Hz)^{-1/2}$ and extrapolating for the full 42 h data set (comprising more than 151 000 data points), we find a relative statistical uncertainty of $5 \times 10^{-16}$ (or 0.1 Hz at 194 THz).

To estimate the uncertainty of the synthesized absolute frequency we also include the systematic uncertainty contributions: that of the caesium fountain as $1.2 \times 10^{-15}$ [26] and that of the $^{171}Yb^+$ frequency standard [27] as $1.5 \times 10^{-15}$. We thus obtain a total relative uncertainty of $2 \times 10^{-15}$. To our knowledge, this is to date the most accurate realization of an optical frequency or wavelength in the optical telecommunications window at 1.5 μm.

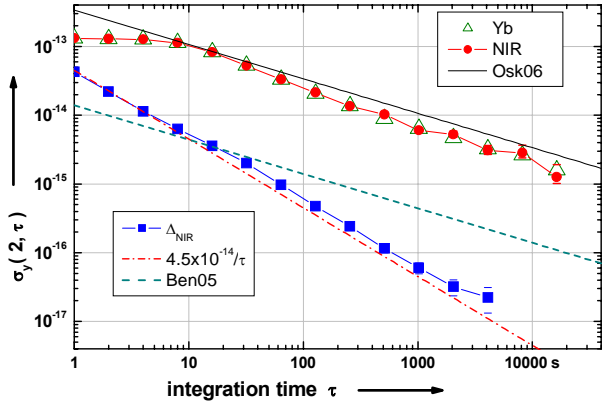

**Figure 2:** Fractional frequency instability (i) of Yb$^+$-laser compared to caesium fountain (green triangles); (ii) of synthesised frequency near 194 THz with the NIR fibre laser compared to caesium fountain (red full circles); (iii) reported by Oskay *et al.* 2006 for Hg$^+$ compared to caesium fountain, using "total deviation" (black continuous line); (iv) of $\Delta_{NIR}$, the frequency difference between two simultaneous frequency comb measurements using independent femtosecond fibre combs (blue full squares) to measure the synthesised NIR frequency versus the caesium fountain (v) a guide to the eye of $1/\tau$ dependence for $\Delta_{NIR}$ (dash-dot red line) (vi) reported by Benkler *et al.* 2005 to represent an upper limit for fluctuations introduced by fibre frequency combs (blue dashed line).

We assess our measurement set-up by measuring the slave laser (NIR) simultaneously with two independent frequency combs versus the microwave frequency derived from CSF1. We thus detect noise contributions e.g. from the beat-detection, frequency counters and from the fibre frequency combs themselves, whereas frequency fluctuations of the NIR laser or of the microwave reference frequency are common mode to both measurement systems, and are thus suppressed.

The Allan standard deviation of the frequency difference $\Delta_{NIR}$ (t) = $\nu_{L2}$(comb2,t) - $\nu_{L2}$(comb1,t) in Fig. 2 (full squares) shows an instability of $4\times10^{-14}$ (1 s) with nearly $1/\tau$ behaviour, reaching $2 \times 10^{-17}$ at 4000 s. This gives an upper limit on frequency fluctuations attributable to the fibre-based frequency combs; for times >20 s our data lie below results obtained by Benkler *et al.* (shown as broken line in fig. 2). We attribute the measured instability mainly to phase noise contributions from the detection and signal processing of $f_{rep}$, as we detect a high harmonic of $f_{rep}$. This involves a commercial multiplier which multiplies the output of a 100 MHz quartz oscillator to 11.4 GHz and appears to degrade its stability.

For the uncertainty contribution of our measurement set-up, we obtain a value of $8 \times 10^{-18}$ for the full data set of $\Delta_{NIR}$ with 32 298 points, if we assume $\tau^{-1/2}$ behaviour between 4000 s and 32 000 s. Taking a more conservative estimate for the uncertainty by using the value of ADEV at 4000 s, the measurement set-up still only contributes $2 \times 10^{-17}$ to the total uncertainty. The averaged frequencies measured with the two frequency combs agreed within this uncertainty: the difference of the mean was 1.6 mHz (corresponding to a fractional difference of $8 \times 10^{-18}$).

The long term instability of the Yb$^+$ optical frequency standard is well documented to be $9 \times 10^{-15}(\tau$ Hz$)^{-1/2}$ for $\tau > 100$ s [28]; by a direct comparison with another optical frequency standard it has been confirmed to be below $2 \times 10^{-15}$ for 1 s < $\tau$ < 100 s.

Having accounted for the instability of our measurement system, and of the optical frequency standard, the only remaining contributor to the instability of the absolute frequency measurement is the rf-reference frequency derived from the Cs fountain clock: this dominates the measurement instability. To overcome this limitation, we performed a direct measurement of the ratio of two optical frequencies.

### 3.2. Measurement of an optical frequency ratio: measuring the synthesised optical frequency versus the optical frequency of the master laser.

Figure 3 shows the results of long-term ratio measurements of two optical frequencies. One frequency comb, comb1, was used for synthesising the optical frequency of the slave laser from the master laser: as described in Sec. 2 we generate an analogue signal $\nu_C$ that can be considered as a beat note between the fixed master $\nu_{L1}$ and the tuneable slave laser $\nu_{L2}$. For practical reasons we divided the signal by 4, thus the virtual carrier frequency corresponds to $\nu_{L2} \times \frac{m_1/2}{m_2} \times \frac{1}{4} \approx 43\,\text{THz}$.

The *same* beat signals used for generating $\nu_C$ and stabilising the slave laser were also counted and used to calculate the frequency ratio $R_1 = \nu_{L2}/\nu_{L1}$ (comb1,t) (red circles in Fig. 3). This frequency ratio therefore represents an "in-loop" measurement. Its relative instability is $8 \times 10^{-16}$ at one second, decreasing as $1/\tau$ between 0.05 s and 500 s, and eventually falling below $10^{-18}$. Note the relative instability of this ratio lies more than a factor 50 below that of $\Delta_{NIR}$.

The second frequency comb, comb2, delivers an independent, out-of-loop measurement of the optical frequency ratio, $R_2 = \nu_{L2}/\nu_{L1}$ (comb2, t). The out-of-loop measurement taken in 2005 (open green triangles in Fig. 3) was reproduced in 2006 (open blue squares in Fig. 3): we obtained a relative instability below $2 \times 10^{-15}$ (1 s) and $5 \times 10^{-18}$ (8000 s) relative to the master laser.

We attribute the constant level from 1 s to 10 s, and the difference in instability between out-of-loop and in-loop signal, primarily to the exposed beam paths and fibre leads in our set-up, which are subject to environmental perturbations. This was tested by duplicating the beat detection of beat $\Delta_2(t)$ between the NIR-laser and frequency comb1, using separate fibre components (fibre leads and beam combiners) and two photo-detectors. We thus observed fluctuations of $5 \times 10^{-16}$ between 1 s and 10 s, which drop to $1 \times 10^{-17}$ at 1000 s.

Residual noise in detection and signal processing was found to contribute about $8 \times 10^{-17}$ in relative instability at 1s, falling off as $1/\tau$; the counter resolution would limit us to a relative instability of approximately $3 \times 10^{-17}/(\tau\,\text{Hz})$.

In order to analyse the in-loop and out-of-loop transfer beats further, we have measured the phase noise of these signals at higher Fourier frequencies by means of a spectrum analyser; here we used the analogue signals $\nu_c$ generated for each comb separately. Figure 4 shows the results for two different settings of the total gain of the phase locked loop.

For very low gain the slave laser is essentially free running (red curve) and shows white frequency noise corresponding to a line width of about 2 kHz, which is indicated by the black dashed line. Increasing the gain to its optimal value, the frequency fluctuations are significantly reduced (black squares) to a level which is limited by the bandwidth of the piezo actuator of the slave laser.

Approximating the measured phase noise by a constant value of -60 dBc for 3 Hz $< f <$ 10 kHz one calculates an Allan standard deviation $\sigma_y(\tau) = 9 \times 10^{-18} \cdot (f_h^{1/2}/(\tau\,\text{Hz}))$, where $f_h$ determines the effective low pass filter of the loop [29]. For $f_h = 10$ kHz the Allan standard deviation of $\sigma_y(\tau) = 9 \times 10^{-16}/(\tau\,\text{Hz})$ is in good agreement with the results obtained by our frequency counting system.

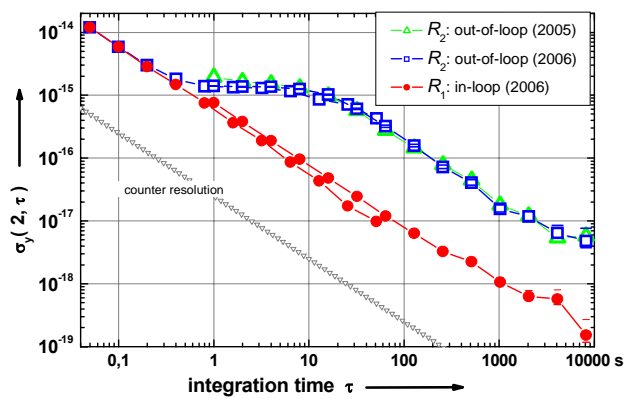

**Figure 3:** Fractional frequency instabilities of the NIR fibre laser calculated by relating its frequency to the Yb$^+$ frequency standard as described by Eq. 5. Measurements were taken in 2005 and 2006; short-term stability (0.05 s to 80 s) and long-term stability (longer than 1 s) were recorded in separate runs.

For Fourier frequencies between 1 Hz and 10 Hz the remaining out-of loop noise starts to increase again and deviates from the in-loop noise shown as grey circles. As discussed above, we attribute this increase primarily to the exposed fibre leads in our set-up. Frequencies below 1 Hz cannot be analyzed as the filter bandwidth of our rf-spectrum analyser is limited to 1 Hz.

Having demonstrated that the residual phase noise of the locked slave laser is below that of the master laser, we conclude that the stability of the master has been transferred to the slave successfully.

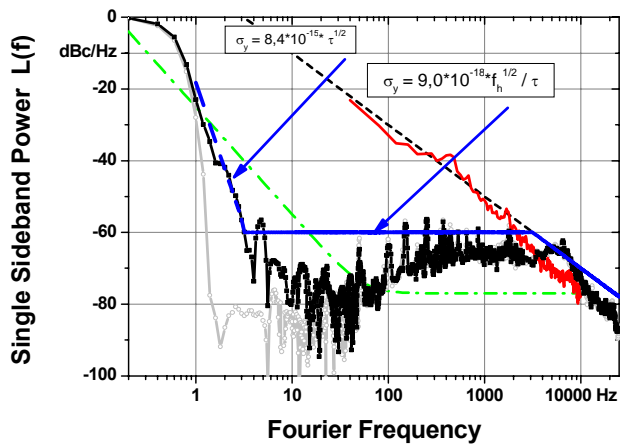

**Figure 4:** Phase noise spectral density of the transfer beat at a virtual carrier frequency of 43 THz derived with comb1 (in loop, grey circles) and comb2 (out-of-loop) for different settings of the gain of the phase-locked loop (PLL, in Fig. 1) and approximation by a constant phase noise level of -60 dBc / Hz for 3 Hz < f < 3 kHz (blue line). For comparison, the phase noise of the Yb master laser as estimated from previous frequency measurements is shown as dashed-dotted green line. The resolution bandwidth of the spectrum analyser was set to 1 Hz.

The *difference* of two independent frequency measurements describes the contribution to the uncertainty of the measuring equipment itself. A direct comparison of the measurement of absolute frequency and frequency ratio is shown in Figure 5: we plot here the absolute frequency difference $\Delta_{NIR}(t)$ and the difference of frequency ratios (re-scaled to 194 THz, the frequency of the NIR laser) as obtained with the two combs. It is immediately apparent that much better short term stability is obtained when measuring the ratio of two optical frequencies. Noise contributions from the detection and signal processing of the repetition rate $f_{rep}$, which probably dominate $\Delta_{NIR}$ (see 3.1), enter with a large multiplication factor; this is not the case for noise involved in the generation and detection of the optical beat signals $v_{ceo}$, $\Delta_1$, and $\Delta_2$, which enter the calculation of $\Delta_R$.

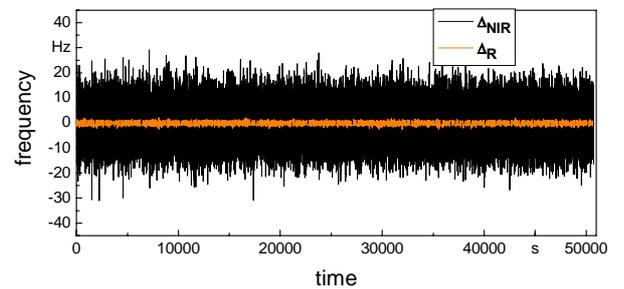

**Figure 5:** Time records of the frequency *difference* of two independent frequency comb measurements of the NIR-laser frequency (i) in black: $\Delta_{NIR}(t)$ (ii) in orange: $\Delta_R(t)$, scaled to the carrier frequency of the NIR laser at 194 THz.

For the measurement of the optical frequency ratio, the instability of the out-of-loop measurement of comb2 gives an upper limit for the noise contributions of the measurement system, including the two frequency combs: involved in this measurement is the synthesis of the slave laser frequency via comb1, the distribution of the slave laser radiation with standard telecom fibres and the ratio measurement with comb2.

The mean values for the frequency ratio as measured by comb1 and by comb2 differed by $6 \times 10^{-18}$: this corresponds to an absolute frequency difference of 1.2 mHz at 200 THz. From the ADEV data, we deduce an uncertainty of $2 \times 10^{-18}$ for the full dataset of more than 50 000 data points; the more conservative estimate, using the last ADEV data point itself, gives an uncertainty of $5 \times 10^{-18}$.

This value gives the precision we achieved for the optical frequency synthesis, by measuring the slave laser frequency in units of the frequency of the master laser.

## 4. CONCLUSIONS

We have shown that an ultra-stable and precisely known optical frequency can be synthesised starting from an ultra-stable master laser. Our synthesised frequency at 194 THz is the most precise realisation of an optical frequency (uncertainty of $2 \times 10^{-15}$) in the long-distance optical telecommunications window to date; we expect that any wavelength within the output spectrum of a frequency comb could be synthesised in this way. The determination of the absolute frequency was limited by the uncertainty of the primary frequency standard itself; relative to the master laser, we obtained an instability below $2 \times 10^{-15}$ (1 s) and $5 \times 10^{-18}$ (8000 s). Such frequency synthesis opens new paths for precision spectroscopy experiments: as an example, one dedicated, optimised and environmentally shielded ultra-stable master laser could be used to distribute any number of ultra-stable frequencies, instead of having to maintain a multitude of such ultra-stable laser systems.

All our measurements were simultaneously recorded using two independent fibre-laser based femtosecond comb generators of fundamentally different design. We observed an agreement better than $8 \times 10^{-18}$ when measuring optical frequencies against a RF reference, and better than $6 \times 10^{-18}$ when measuring the ratio of two optical frequencies. The low instability and high accuracy observed open the possibility of characterising laser sources relative to a master laser in a different spectral region. The beat notes involved in the measurements covered a broad spectral range (modes near 1 μm and 2 μm for $\nu_{ceo}$, 1742 nm for $\nu_{L1}$, 1543 nm for $\nu_{L2}$) of both frequency combs. We found no indication that the instability in our frequency measurements is caused by fluctuations of the frequency combs themselves.

The overall experiment involved converting the stable frequency of a local optical frequency standard operating at 344 THz to an optical carrier frequency of 194 THz using one frequency comb, transmitting this laser light through standard telecommunication fibre and telecom components to a second frequency comb, where we then measured the frequency of an optical frequency standard in units of the transmitted light frequency. We have thus presented the ingredients and current measurement floor for a remote measurement of optical frequencies by using a carrier in the optical telecommunication window.

**Acknowledgement**

This work was supported by the Deutsche Forschungsgemeinschaft (DFG) under SFB 407. The authors would like to thank Christian Tamm for providing an optical frequency reference, Uwe Sterr for useful discussions and Christian Lisdat for proof-reading the manuscript.